\documentclass[12pt]{article}
\usepackage{tikz}
\usepackage{color}
\usepackage[numbers]{natbib}
\usepackage{float}
\usepackage[mathscr]{euscript}
\usepackage{comment}

\usepackage{geometry} 
\usepackage[applemac]{inputenc} 
\usepackage{textcomp} 
\usepackage{graphicx}  
\usepackage{amsmath,amssymb}  
\usepackage{bm}  

\usepackage{amsfonts}

\def\dim{\mathop{\hbox{dim}}}

\newtheorem{teor}{Theorem}[section]

\newtheorem{remar}[teor]{Remark}

\newtheorem{lemma}[teor]{Lemma}

\newcommand{\fdim}{\hspace*{\fill}$\Box$}
\newcommand{\dimostr}{{\bf Proof: }}
\newcommand{\real}{\mathbb{R}}

\newcommand*{\rom}[1]{\expandafter\@slowromancap\romannumeral #1@}
\newcommand{\RNum}[1]{\uppercase\expandafter{\romannumeral #1\relax}}

\begin{document}
\title{Minimal entropy and uniqueness of price equilibria in a pure exchange economy
\thanks{The first author was  supported  by INdAM. GNSAGA - Gruppo Nazionale per le Strutture Algebriche, Geometriche e le loro Applicazioni, by KASBA  Funded by Regione Autonoma della Sardegna and by STAGE- Funded by Fondazione di Sardegna.The second author was supported by GROVID and by STAGE, funded by Fondazione di Sardegna and Regione Autonoma della Sardegna.}
}

\author{Andrea Loi\thanks{Dipartimento di Matematica e Informatica, email:
loi@unica.it.  
}\\
\and
Stefano Matta\thanks{Correspondence to
S. Matta, Dipartimento di Economia, University of Cagliari, 
viale S. Ignazio 17, 09123 Cagliari, Italy,
tel. 00390706753340, Fax  0039070660929
email: smatta@unica.it}\\
\and\centerline{(University of Cagliari)}}

\maketitle

\vspace{0.4in}

\noindent\textbf{Abstract}:

\noindent 
We introduce uncertainty into a pure exchange economy and establish a connection between Shannon's differential entropy 
and uniqueness of price equilibria.
The following conjecture is proposed under the assumption of a uniform probability distribution:  entropy is minimal
if and only if the price is unique for every economy.
We show the validity of this conjecture  for an arbitrary number of goods and two consumers
and, under certain conditions, for  an arbitrary number of consumers and two goods. 
\vskip 0.3cm

\vspace{0.3in}

\noindent\textbf{Keywords:}  Entropy, uniqueness of equilibrium, price multiplicity, equilibrium manifold, minimal submanifold.
\vspace{0.3in}

\noindent\textbf{JEL Classification:} D50, D51, D80.
\newpage

\section{Introduction}\label{intro}

\noindent 

In a pure exchange economy let us denote by $x=(p,\omega)$ an initial allocation $\omega$ and 
its supporting equilibrium price vector $p$. Suppose that $x$ is slightly perturbed by
exogenous, i.e. shocks, or endogenous factors,
e.g. the uncertainty related to the effects of Safra's competitive manipulation \cite{saf}.
The result of this perturbation is a new allocation and equilibrium price vector, $x'$, belonging
to a neighborhood $N$ of $x$. 
We can represent this process as a probability
model, where the random variable ranges in the set $N$. 
Observe that $N$ is not an Euclidean space. It belongs to
a space of endowments and prices and consists of points such that
aggregate excess demand function is equal to zero.
Under standard smooth assumptions and in a fixed total resource setting, 
$N$ becomes a submanifold with boundary  of a manifold
called the {\em equilibrium manifold}, denoted by $E(r)$ 
which in turn is  a smooth submanifold of $S\times \Omega (r)$, where $S$ is the space of prices and
$\Omega (r)$ the space of economies
 (see the seminal work by Balasko \cite{balib} and also Section \ref{ecoprel}).

Thus $E(r)$ can be equipped with a natural  measure, namely the Riemannian volume form
$dM_g$ associated to the Riemannian metric $g$
induced by the flat metric of its ambient space $S\times \Omega(r)$
(see e.g. \cite{lmgc}).
The probability that $x\in E(r)$ belongs to $N$ is  
$$Pr(x\in N)=\int_Np(x)dM_g(x),$$
where $p$  is a given  probability density on $E(r)$
(the reader is referred to \cite{pennec} for a geometric approach to probability theory
on Riemannian manifolds).
Moreover, following Shannon \citep{sw} (see also  \cite{ct})
in this framework  we define  the {\em differential entropy of $N$} as
$$H(N)=-\int_{N} p(x)log (p(x))dM_g(x).$$  
Obviously when $E(r)$ is an Euclidean flat space
then one can take $dM_g$ equals to the Lebesgue measure 
and $H(N)$ is the differential entropy defined in \citep{sw}.

Since entropy is a measure of missing information it is natural to investigate under which conditions it is minimized.
Therefore we provide he following: 
\vskip 0.3cm
\noindent
{\bf Definition (MEP)} {\em The  equilibrium manifold satisfies the minimal entropy  property
(MEP) if for every economy $x$ belonging to $E(r)$, there exists a neighborhood $N$ of $x$ in $E(r)$
such that 
$H(N)\leq H(\tilde N)$, where
$\tilde N$ is any other submanifold of $S\times\Omega(r)$  containing $x$ which has 
the same boundary of $N$ and whose volume structure, in the same way as $N$,
is induced by the  flat metric of the ambient space $S\times \Omega (r)$.}
\vskip 0.3cm

It would be interesting (and challenging) to study how the choice of different probability
distributions affects the economic properties implied by (MEP). This issue, which deserves further analysis,
is beyond the scope of this paper.

On the other hand, it is natural to restrict
to the case of uniform distribution, 
namely when the probability density function is given by
$$p_N=\frac{\chi_n}{V(N)},$$
where $V(N)=\int_NdM_g(x)$ is the volume of $N$
and $\chi_N:E(r)\rightarrow \{0, 1\}$ is the characteristic function supported on $N$.
Under this assumption 
\begin{equation}\label{de}
H(N)=-\int_N\frac{1}{V(N)}log( \frac{1}{V(N)})dM(x)=log(V(N)),
\end{equation}

and so by the  increasing property of the logarithm we  deduce that  in the case of uniform 
distribution the  (MEP) is equivalent 
to the following

\vskip 0.3cm
\noindent
{\bf Definition (MVP)} {\em The  equilibrium manifold satisfies the minimal volume  property
(MVP) if for every economy $x$ belonging to $E(r)$, there exists a neighborhood $N$ of $x$ in $E(r)$
such that 
$V(N)\leq V(\tilde N)$, where
$\tilde N$ is any other submanifold of $S\times\Omega(r)$  containing $x$ which has 
the same boundary of $N$.}
\vskip 0.3cm

Now the (MVP) for $E(r)$ can be translated into the language  of  differential geometry: the equilibrium manifold $E(r)$ is a {\em stable minimal submanifold} of $S\times\Omega (r)$, i.e.  a local minimum of the volume functional. In particular  $E(r)$ is a critical point of the volume functional, 
namely  $E(r)$ is a {\em minimal submanifold}\footnote{Throughout this paper we will content ourselves with this definition since in the proof of our main results  we are not using the differential geometric machinery of the theory of minimal submanifolds.
 The interested reader is  referred to \cite{sim} for more details and material on minimal submanifolds.
 The simplest examples of minimal submanifolds arise when $n=1$: in this case they are simply geodesics of
 the ambient space. 
In higher dimensions every totally geodesic submanifold is a minimal submanifold (cf. also \cite{geodequ} for some properties of  geodesics and totally geodesic submanifolds of the equlibrium manifold).
Nevertheless, there exist a lot of interesting minimal submanifolds (see  \cite{sim} or Section \ref{main} below).}of $S\times \Omega (r)$.

Observe now that according to  Theorem \ref{flatman} below,  if for every economy there is uniqueness of equilibrium, the equilibrium manifold
is \lq\lq flat'' (and hence minimal): i.e., (global) uniqueness implies (MVP).
Here  we explore the reverse of this  implication: if there is price multiplicity, can (MVP) holds true? 
In other words, does (MVP)  implies uniqueness?
This is not a trivial issue: in fact the equilibrium manifold can almost arbitrarily be twisted for an appropriate preference profile\footnote{Even if the equilibrium manifold  $E(r)$ is unknotted in its ambient space \citep{demger}.}.
Hence one could expect to find an utility profile which gives rise to multiplicity and minimality. 
Actually, we believe that exactly the opposite is true. Indeed we address the following:

\vskip 0.3cm
\noindent
{\bf Conjecture:} {\em Under the assumption of uniform distribution  the equilibrium manifold
satisfies (MEP)  if and only if  the equilibrium price is unique.} 
\vskip 0.3cm

In other words, we believe that an utility profile which minimizes entropy (and hence volume) with uniform distribution  is incompatible with price multiplicity.

In this paper we show the validity of this conjecture in the case of an arbitrary number of goods and two consumers (Theorem \ref{teorm=2})
and in the case of an arbitrary number of consumers and two goods (Theorem \ref{teorl=2}) under the additional assumption that the normal vector field of $E(r)$ is constant outside a compact subset of the ambient space. 
The proof of Theorem \ref{teorm=2} strongly relies on geometric and economic properties:
the classification of ruled minimal submanifolds of the Euclidean space, the bundle structure of the equilibrium manifold and the positiveness of prices.
On the other hand, the proof of Theorem \ref{teorl=2} combines  deep geometric results relating the topology of a minimal submanifold of the Euclidean space  with the fact that $E(r)$ is globally diffeomorphic to an Euclidean space.

It is worth noticing that the choice of a metric depends on the analysis.
In \cite{lmmeas} the metric on the equilibrium manifold was chosen to deal with asymptotic properties
related to economies with an arbitrarily large number of equilibria.
In \cite{lmgc} the metric used was a tool to explore geometric properties which are intrinsic, 
i.e. they do not depend on the ambient space.
But the purpose, and the approach, of the present work is entirely different and this affects the choice of
the metric used.

We believe that this information-theoretic and geometric approach 
can be further extended in different directions.
Following the seminal contribution by \cite{theil} (see also \cite{cow,cow2} and \cite{maa}),
an entropy-based metric could be developed in order to compute geodesics representing 
redistributive policies. Another direction of research (see \cite{saf} and \cite{goma}) 
is to analyze the extrinsic uncertainty in $N\subset E(r)$  caused by coalitional manipulation of endowments.
This approach could provide new insights into the understanding of coalition formation and sunspot equilibria.
Finally, due to the economic relevance of the shape of the equilibrium manifold,
it can be worth investigating the connection between its shape and the primitives of the model, an issue still largely unexplored.
This local-global view can hopefully
lead to new perspectives on issues like uniqueness and stability (see \cite{kehoe,msc} for a survey).

This paper is organized as follows. 
Section \ref{ecoprel} recalls notations, definitions and the existing results which will be used to prove our main results. 
Section \ref{main} and Section \ref{main2} prove our main results, Theorem \ref{teorm=2} and Theorem \ref{teorl=2}.

\section{Definitions}\label{ecoprel}

The economic setup is represented by a pure exchange smooth economy with $L$ goods and $M$ consumers
under the standard smooth assumptions (see \cite[Chapter 2]{balib}).
The set of normalized prices is defined by
$$S=\{p=(p_1,\dots p_L)\in\real^L\ |\ p_l>0, l=1,\ldots,L,\ p_L=1\}$$
and the set $\Omega =({\real}^{L})^M$ denotes the space of endowments 
$\omega =(\omega_1, \dots ,\omega_M)$, $\omega_i\in {\real}^L$. 
The {\em equilibrium manifold} $E$ is the set of the pairs
$(p, \omega)\in S\times \Omega$, which satisfy the equality:
\begin{equation}\label{E}
\sum_{i=1}^{M}f_i(p, p\cdot \omega_i)=\sum_{i=1}^{M}\omega_i,
\end{equation}

\noindent where $f_i(p, w_i)$ is consumer's $i$ demand.

\noindent By \cite[Lemma 3.2.1]{balib},
$E$ is a (closed) smooth submanifold
of $S\times \Omega$, globally diffeomorphic to 
$S\times\real^M\times\real^{(L-1)(M-1)}={\real}^{LM}$,
i.e. $\phi_{|E}\cong \real^{LM}$,  where the smooth mapping
$$\phi:S\times \Omega\to S\times\real^M\times\real^{(L-1)(M-1)}$$ 
is defined by $$(p,\omega_1\ldots,\omega_M)\mapsto (p,p\cdot\omega_1,\ldots ,p\cdot\omega_M,\bar\omega_1,\dots,\bar\omega_{M-1}),$$
where $\bar\omega_i$ denotes the first $L-1$ components of $\omega_i$,
for $i=1,\ldots,M-1$.

\vskip 1cm
\noindent We also introduce the following two subsets of $E$:
\begin{itemize}
\item the set of {\em no-trade equilibria} $T=\{(p,\omega)\in E|\, f_i(p,p\cdot\omega_i)=\omega_i,\, i=1,\ldots, M\}$;
\item {\em the fiber} associated with 
$(p,w_1,\ldots,w_M)\in S\times\real^M$,
which is defined as the set of pairs 
$(p,\omega)\in S\times\Omega$ such that:
\begin{itemize}
\item $p\cdot\omega_i=w_i$ for $i=1,\ldots, M$;
\item $\sum_i\omega_i=\sum_if_i(p,w_i)$.
\end{itemize}
\end{itemize}

\noindent By defining the two smooth maps
$$f:S\times\real^M\to S\times\real^{LM},$$ 
where $f(p,w_1,\ldots, w_M)=(p,f_1(p,w_1),\ldots, f_M(p,w_M))$,
and 
$$\phi_{Fiber}:E\to S\times\real^M,$$ 
where
$\phi_{Fiber}(p,\omega_1,\ldots,\omega_M)=(p,p\cdot\omega_1,\ldots,p\cdot\omega_M)$,  since $f(S\times\real^M)=T \subset E$ and $\phi_{Fiber}\circ f$
is the identity mapping, by applying \cite[Lemma 3.2.1]{balib},
Balasko shows \citep[Proposition 3.3.2]{balib} that T is a smooth submanifold of $E$ diffeomorphic to $S\times\real^M$. 

By construction, every fiber associated with $(p,w_1,\ldots,w_M)$  is a subset of $E$ which is the inverse image of $(p,w_1,\ldots,w_M)$ via the mapping $\phi_{Fiber}$. It is intuitively clear that while holding  
$(p,w_1,\ldots,w_M)$ fixed
and letting $\omega$ varying along the fiber, there are not
any nonlinearities which may arise from the aggregate demand.
In fact the fiber is a linear submanifold of $E$ of dimension $(L-1)(M-1)$
\citep[Proposition 3.4.2]{balib}.

Since every fiber contains only one no-trade equilibrium
\citep[Proposition 3.4.3]{balib},  
the equilibrium manifold $E$ can be thought
as a disjoint union of fibers parametrized by the 
no-trade equilibria $T$ via the mapping $\phi_{|E}:E\to S\times\real^M\times\real^{(L-1)(M-1)}$: for a fixed $(p,w_1,\ldots,w_M)\in S\times\real^M$, each fiber is parametrized by $\bar\omega_1,\ldots,\bar\omega_{M-1}$. By letting 
$(p,w_1,\ldots,w_M)$ varying in $S\times\real^M$, we obtain
the {\em bundle structure} of the equilibrium manifold.

\vskip 0.5cm
\noindent If total resources are fixed,
the equilibrium manifold is defined as

\begin{equation}\label{Er}
E(r)=\{(p, \omega)\in S\times \Omega (r)\ | \sum_{i=1}^Mf_i(p, p\cdot \omega_i)=r\},
\end{equation}

\noindent where $r\in {\real}^L$ is the vector that represents the total resources of the economy and $\Omega (r)=\{\omega\in {\real}^{LM}\ |\  \sum_{i=1}^M\omega_i=r\}$.

\noindent Let 
\begin{equation}\label{Br}
B(r)=\{(p,w_1,\ldots,w_M)\in S\times\real^M|\, \sum_{i=1}^Mf_i(p,w_i)=r\}
\end{equation}

\noindent be the set of {\em price-income equilibria}
(see \cite[Definition 5.1.1]{balib}).
\noindent $B(r)$ is a submanifold of $S\times\real^M$
diffeomorphic to $\real^{M-1}$ \citep[Corollary 5.2.4]{balib}
through the map
${ \theta}:S\times\real^M\to\real^L\times\real^{M-1}$,
defined by

\begin{equation}\label{theta}
(p,w)\mapsto (\sum_if_i(p,w_i),u_1(f_1(p,w_1),\ldots, u_{M-1}(f_{M-1}(p,w_{M-1})).
\end{equation}

\noindent The equilibrium manifold $E(r)$
is a submanifold of $S\times\Omega(r)$ diffeomorphic to $\real^{L(M-1)}$
\citep[Corollary 5.2.5]{balib}

\begin{equation}\label{mappar}
\phi(E(r))=B(r)\times\real^{(L-1)(M-1)}.
\end{equation}

\noindent Moreover we can define and 
$T(r)=T\cap S\times\Omega(r)$.
By construction, even in a fixed total resource setting,
the equilibrium manifold preserves its bundle structure property and, hence,
$E(r)$ can be written as the disjoint union 

\begin{equation}\label{bundlestr}
E(r)=\sqcup_{x\in T(r)}F_x, 
\end{equation}
where $F_x$ is an $(L-1)(M-1)$-affine subspace of
 $\real^{L(M-1)}$.

\noindent 

Let $t=(t_1,\ldots,t_{l-1})$, $\bar\omega_j=(\omega_1^1,\ldots,\omega_1^{l-1})$
and $p(t)=(p_1(t),\ldots,p_{l-1}(t)$.
Following \cite{balib} and \cite{lmgc},
we can parametrize $B(r)$ via the map:
\begin{equation}\label{parbr}
\varphi:\real^{M-1}\to B(r),\,t\to (p(t),w_1(t)\ldots,w_{m-1}(t))
\end{equation}
and  $E(r)$ via the map:

\begin{equation}\label{parer}
\Phi:\real^{L(M-1)}\to E(r), 
\end{equation}
$$(t,\omega_1^1,\ldots,\omega_{M-1}^1,\ldots,\omega_1^1,\ldots,\omega_{M-1}^{L-1})
\mapsto (p(t),\bar\omega_1,w_1(t)-p(t)\cdot\bar\omega_1,\ldots,w_{M-1}(t)-p(t)\cdot\bar\omega_{M-1}) 
$$

We end this section with the following result due to Balasko, deeply related to the issue raised in this paper.

\begin{teor}{\rm\citep[p. 188 Theorem 7.3.9 part (2)]{balib}}\label{flatman}
If for every $\omega\in\Omega(r)$ there is uniqueness of equilibrium,
the equilibrium correspondence is constant: The equilibrium price
vector $p$ associated with $\omega$ does not depend on $\omega$.
\end{teor}

\begin{remar}\rm\label{remarbal}
Hence (global) uniqueness implies  (MEP)  for $E(r)$ under the assumption  of a  uniform distribution.
This theorem will be used 
to prove the \lq\lq only if" part of Theorem \ref{teorm=2} and Theorem \ref{teorl=2}.
\end{remar}

 \section {The case $M=2$}\label{main}

In this section we prove the following:

\begin{teor}\label{teorm=2}
Let $M=2$ and assume a uniform distribution.
Then  $E(r)$ satisfies the (MEP) if and only if the price is unique.
\end{teor}

Before proving the theorem  we need some definitions.

\begin{itemize}

\item a submanifold $\mathscr M^n\subset\real^{n+k}$
is said to be ruled if $\mathscr M^n$ is foliated by affine subspaces of dimension $n-1$ in $\real^{n+k}$.

\item a  {\em generalized helicoid} is the ruled  submanifold $\mathscr M^n(a_1, \dots ,a_k, b)\subset\real^{n+k}$,   
$k\leq n$, admitting the following parametrization:
\end{itemize}
$$(s,t_1,\ldots, t_{n-1})\mapsto(t_1\cos(a_1s),t_1\sin(a_1s),\ldots,t_k\cos(a_ks),t_k\sin(a_ks), t_{k+1},\ldots,t_{n-1},bs)),$$

where $a_j,\, j=1,\ldots, k$, and $b$ are real numbers (we are not escluding that one of these coefficients could vanish and the generalized helicoid 
becomes an affine subspace).

The key ingredient in the proof of Theorem \ref{teorm=2} is the 
following classification result on  ruled minimal submanifolds of the Euclidean space. We refer the reader to \cite[Section 1]{dil} and references therein (in particular \cite{lum} for a proof).

\begin{teor}[\cite{lum}]\label{lum}
A minimal ruled submanifold $\mathscr M^n\subset\real^{n+k}$ is, up to rigid motions\footnote{A rigid motion of the Euclidean space $\real^{l}$ is an isometry of $\real^{l}$ given by  the composition of an orthogonal  $l\times l$ matrix and a translation by some vector $v\in \real^l$.} of $\real^{n+k}$, a generalized helicoid.
\end{teor}

We need also the following simple but fundamental fact:

\begin{lemma}\label{hcont}
Let $\mathscr M^n(a_1, \dots ,a_k, b)\subset\real^{n+k}$  be a generalized helicoid such that 
$b\cdot\prod_{i=1}^ka_i\neq 0$. Then $\mathscr M^n$ intersects any affine hyperplane of $\real^{n+k}$.
\end{lemma}
\noindent\dimostr
In cartesian coordinates
$x_1, y_1, \dots , x_k, y_k, x_{k+1}, \dots , x_{n-1}, x_n$
an hyperplane of $\real^{n+k}$ has equation:
$$\alpha_1x_1+\beta_1y_1+\cdots +\alpha_kx_k+\beta_ky_k+\alpha_{k+1}x_{k+1}+\cdots +\alpha_{n-1}x_{n-1}+\alpha_n x_n=\delta,$$
where $\alpha_i, \beta_i, i=1, \dots k$, $\alpha_j, j=k+1, \dots n$ and  $\delta$ are real numbers such that   
$$\sum_{i=1}^k(\alpha_i^2+\beta_i^2)+\sum_{j=1}^n\alpha_j^2\neq 0.$$

On the other hand the following equation
represents  the condition to be satisfied 
for a point of the hyperplane to belong to the generalized helicoid:
$$\sum_{i=1}^kt_i(\alpha_i\cos(a_is)+\beta_i\sin(a_is))+\sum_{j=k+1}^{n-1}\alpha_jt_j+\alpha_nbs=\delta.$$
Since one can always find a pair $(s_0, t_0)$ satisfying the previous equation, the lemma is proved. 
\fdim

\vskip 0.3cm
\noindent
{\bf Proof of Theorem \ref{teorm=2}:}
Since  (MEP) is equivalent to  (MVP),
$E(r)$ is a minimal submanifold of $S\times \Omega (r)$.
Since  $M=2$, by the bundle structure property 
(see (\ref{bundlestr}) above)
$E(r)$ is a ruled submanifold  in $\real^{2L-1}$.
By Theorem \ref{lum}, $E(r)$ is (up to rigid motions) a  generalized helicoid.
If some $a_i$ or $b$ are equal to zero then 
$E(r)$ is an hyperplane and, by Theorem \ref{flatman},  the price is unique. Otherwise if $b\cdot\prod_ia_i\neq 0$, by combining  Lemma \ref{hcont} 
with the fact that  $E(r)$ is contained in the open set of $\real^{2L-1}$ consisting of those points with $p>0$ ($p$ being the price)
one deduces that $E(r)$ is  an affine hyperplane and so the price is unique.
The \lq\lq only if" part follows by Theorem \ref{flatman} (see Remark \ref{remarbal}). \fdim

\vskip 0.5cm
\noindent 

\begin{remar}\rm
In the previous theorem we use the fact that  $E(r)\subset S\times \Omega (r)$ is a minimal submanifold 
We can prove the same result by only assuming that the no-trade equilibria 
$T(r)$ (which is one dimensional for $M=2$)
is a minimal submanifold of $E(r)$, namely it is a geodesic.
Indeed, by using the diffeomorphism between $T(r)$ and $B(r)$, and the parametrization $\Phi$
of $E(r)$ (see (\ref{parer})), $T(r)$ can be parametrized through $\Phi$ by letting $v=0$:
$$\Phi(t,0)=\gamma(t).$$
Hence, if $T(r)$ is a geodesic in $E(r)$, its acceleration $\gamma''(t)$ is parallel, for every $t$, to the unit normal vector
$N(t)_{|v=0}$ of $E(r)$ or, equivalently, $\gamma''(t)\wedge N(t)_{|v=0}=0$.
We have $\gamma''(t)=\beta''(t)=(\ddot{p},0,\ddot{w}$) and, since $v=0$, 
$\Phi_t\wedge\Phi_v=\dot{\beta}\wedge\delta=(-\dot{w},p\dot{p},\dot{p})$.
Hence $\gamma''(t)\wedge N(t)_{|v=0}=\beta''\wedge(\beta'\wedge\delta)=0$ if and only if 
$$(-p\dot{p}\ddot{w},p\ddot{p}+\dot{w}\ddot{w},p\dot{p}\ddot{p})=(0,0,0).$$

This implies that, for every $t$, $\dot{p}\ddot{p}=0$, i.e. $(\dot{p}\dot{p})'=0$, 
hence $p$ is (constant and) unique. 
\end{remar}


\section{The case $L=2$}\label{main2}
In this section we consider an economy with two goods and an arbitrary number of consumers. In this case the equilibrium manifold is a hypersurface. 
More precisely, the equilibrium manifold $E(r)$ has dimension $\real^{2M-2}$
and the ambient space has dimension $\real^{2M-1}$. So it makes sense to consider
the normal vector field $N$ along $E(r)$, namely for each $x\in E(r)$ we consider a unit vector $N(x)$ parallel
to the affine line $T_xX^\perp$ normal to the tangent space
$T_xX$ of $X$ at $x$.  The smooth map $N:E(r)\rightarrow S^{2M-2}$ which takes $x$ to the point $N(x)$ of the unit sphere $S^{2M-2}\subset \real^{2M-1}$ is called the {\em Gauss map}. Obviously, if the Gauss map is constant then the price is constant and  hence $E(r)$ is an affine hyperplane in $\real^{2M-1}$. 
In the following theorem, which represents the second main result of the paper, we show that the minimality assumption together with
the constancy of the Gauss map outside a compact set imply uniqueness of 
the equilibrium price.

\begin{teor}\label{teorl=2}
Let $L=2$.
Assume that the Gauss map  is constant outside a compact subset of $E(r)$.
Under the assumption of uniform distribution,
$E(r)$ satisfies (MEP) if and only if the price is unique.
\end{teor}

This theorem can be  intepreted by saying that if the equilibrium manifold is minimal and there exists a compact subset $K$ of 
$\real^{2M-1}$ such that  
$(\real^{2M-1}\setminus K)\cap E(r)$ is the union of open subsets  of hyperplanes each parallel to the hyperplane $p=const$, then $E(r)$ is indeed
an hyperplane. 
As a consequence, the usual one-dimension representation
of the equilibrium manifold cannot be minimal (see figure below).

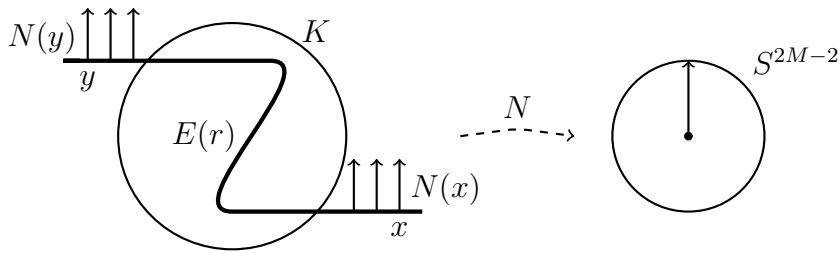
\begin{figure}[H]
\centering
\begin{tikzpicture}
\draw [ultra thick] (0,5) to [out=180,in=180] (2.5,5)
to [out=0,in=180] (2,3) to [out=0,in=180] (4.5,3);
\draw [thick] (2,4) circle [radius=1.5];
\draw [->, thick] (0.7,5) -- (0.7,5.7);
\draw [->, thick] (0.4,5) -- (0.4,5.7);
\draw [->, thick] (0.1,5) -- (0.1,5.7);
\draw [->, thick] (3.6,3) -- (3.6,3.7);
\draw [->, thick] (3.9,3) -- (3.9,3.7);
\draw [->, thick] (4.2,3) -- (4.2,3.7);
\draw [thick] (8,4) circle [radius=1];
\draw [->, thick] (8,4) -- (8,5);
\draw[fill] (8,4) circle  [radius=.05];
\node[left] at (0.1,5.3) {$N(y)$};
\node[below] at (0.1,5) {$y$};
\node[right] at (4.2,3.3) {$N(x)$};
\node[below] at (4.2,3) {$x$};
\node[above] at (5.75,4.1) {$N$};
\node[left] at (2.2,4) {$E(r)$};
\node[above] at (3.1,5.1) {$K$};
\node[right] at (8.7,5) {$S^{2M-2}$};
\draw [->,thick, dashed, rounded corners] (5,4) -- (5.75,4.1) -- (6.5,4);
\end{tikzpicture}
\end{figure}

\vskip 0.3cm
The proof of Theorem \ref{teorl=2} relies on the following theorem obtained in turn by suitably  combining some deep results obtained by Anderson in   \cite{and}.

\begin{teor}\label{lemmone}
Let $\mathscr{M}^n\subset\real^{n+1}$, $n>2$, 
be a minimal hypersurface such that the following conditions are satisfied:
\begin{itemize}
\item [1.]
$\mathscr{M}^n$ has one end; 
\item [2.]
$\mathscr{M}^n$ is a $C^1$-diffeomorphic to 
a compact manifold $\bar{\mathscr{M}}^n$ punctured at a finite number of points
$\{p_i\}$. 
\item [3.]
the Gauss map $N:\mathscr{M}^n\rightarrow S^n$ extends to a $C^1$-map of 
$\bar{\mathscr{M}}^n$.
\end{itemize}
Then $\mathscr{M}^n$ is an affine $n$-plane.
\end{teor}

\begin{remar}\rm
The number of ends of a smooth manifold is a topological invariant which, roughly speaking,  measures the number of connected components
\lq\lq at infinity''. The reader is referred to \cite{and} for a rigorous  definition.
What we are going to use in the proof of Theorem \ref{teorl=2} is that for $n>1$, the  Euclidean space $\real^n$ has only one end. This is because $\real^n\setminus K$ has only one unbounded component for any compact set $K$.
\end{remar}

\vskip 0.3cm
\noindent
{\bf Proof of Theorem \ref{teorl=2}:}
Since  (MEP) is equivalent to  (MVP),
$E(r)$ is a minimal submanifold of $S\times \Omega (r)$.
If $M=2$ we can apply Theorem \ref{teorm=2}.
We  can then assume $M>2$ and so $\dim E(r)=2M-2>2$.
Hence, in order to prove the \lq\lq if'' part it is enough to verify that the three conditions of Theorem \ref{lemmone} are satisfied for $E(r)\subset \real^{2M-1}$.
Condition 1 follows by the previous  remark,
since $E(r)$ is globally diffeomorphic to $\real^{2M-2}$. 
Notice that the unit sphere  $S^{2M-2}$ is the Alexandroff compactification of $E(r)\cong\real^{2M-2}$, namely it can be obtained by adding one point, called $\infty$, to $E(r)$. In other words $E(r)$  is diffeomorphic to the sphere $S^{2M-2}$ punctured to  $\infty$ and so also condition 2 holds true. 
The  assumption that the  Gauss map $N:E(r)\rightarrow S^{2M-2}$ is constant outside a compact set $K$ means that  $N(x)=N_0$, where $N_0$ is a fixed vector in $S^{2M-2}$,
for all $x\in E(r)\setminus K$. Therefore,  one can extend $N$  to a $C^{\infty}$-map  $\hat N: S^{2M-2}\rightarrow S^{2M-2}$  by simply defining 
$\hat N (\infty)=N_0$, and so  also condition 3 is satisfied.
The \lq\lq only if" part follows  by Theorem \ref{flatman} (see Remark \ref{remarbal}).
\fdim

\begin{remar}\rm\label{remarfinal}
Given a submanifold  $\mathscr M^n$ of a Riemannian manifold $\mathscr N^{n+k}$,
one can express the minimality condition in terms of the vanishing of the  mean curvature $H$. If  $k=1$, namely when 
$\mathscr M^n$ is an hypersurface, the minimality condition,  namely $H=0$ ,
is equivalent to the vanishing of the trace of the differential of the Gauss
map (see e.g. \cite{do}). Thus, for $L=2$  one could try to show that minimality of $E(r)$ implies uniqueness of the equiilbrium price  without imposing the extra condition 
on the constancy of the Gauss map outside a compact set (as in Theorem \ref{teorl=2}) by computing the Gauss map through the parametrization (\ref{parer}) above and imposing that the vanishing of the trace of its differential. This gives rise to a complicated PDE equation, which the authors were not able to handle even when $M=3$.
\end{remar}

\small{}

\end{document}